\documentclass[%
 preprint,
superscriptaddress,
 amsmath,amssymb,
 aps, physrev,
]{revtex4-2}

\usepackage{graphicx}
\usepackage{dcolumn}
\usepackage{bm}
\usepackage{hyperref}
\usepackage[english]{babel}

\usepackage[dvipsnames]{xcolor}
\usepackage{soul}          
\usepackage[normalem]{ulem} 

\usepackage{xcolor}

\usepackage[normalem]{ulem} 

\begin{document}


\title{\textbf{Single-Spin Nitrogen-Vacancy Hybrid Magnetometer with Enhanced Static Field Sensitivity} 
}%

\author{Vinaya K. Kavatamane}
\email{vinayakumar.kavatamane@ucalgary.ca}
\altaffiliation{Current address: Institute for Quantum Science and Technology, University of Calgary, Calgary, Alberta T2N 1N4, Canada}

\author{Dewen Duan}
\affiliation{Max Planck Research Group -- Nanoscale Spin Imaging, 
Max Planck Institute for Biophysical Chemistry, 
Am Fassberg 11, 37077 G\"ottingen, Germany}
\altaffiliation{Current address: School of Physics and Electronic Engineering, Sichuan University of Science and Engineering, Yibin 644005, China}

\author{Hadi Zadeh-Haghighi}
\affiliation{QNeura, Calgary, Alberta T2L 0Y2, Canada}

\author{Manh-Huong Phan}
\affiliation{Center for Materials Innovation and Technology, 
VinUniversity, Hanoi 100000, Vietnam}

\author{Gopalakrishnan Balasubramanian}
\altaffiliation{Current address: XeedQ GmbH, Augustusplatz 1--4, 04109 Leipzig, Germany}
\email{gopi@xeedq.com}
\affiliation{Max Planck Research Group -- Nanoscale Spin Imaging, 
Max Planck Institute for Biophysical Chemistry, 
Am Fassberg 11, 37077 G\"ottingen, Germany}

\begin{abstract}
Precision sensing and imaging of weak static magnetic fields are crucial for a variety of emerging nanoscale applications. While nitrogen-vacancy (NV) centers in diamond provide exceptional AC magnetic field sensitivity with nanoscale spatial resolution, their sensitivity to static (DC) magnetic fields is fundamentally limited by the short dephasing time ($\text{T}_2^*$) due to spin-spin interactions. In this work, we present a hybrid sensing approach that integrates a soft ferromagnetic microwire with a single near-surface NV center to amplify its response to external static magnetic fields. This hybrid configuration achieves a DC magnetic field sensitivity of 63\,nT/$\sqrt{\text{Hz}}$ for a single NV center—about 500 times greater than conventional inhomogeneous broadening- or $\text{T}_2^*$-limited magnetometry, with potential for further enhancement. The compact and highly sensitive nature of this sensor opens new opportunities for quantum sensing applications involving the detection of static or slowly varying magnetic fields across diverse scientific and technological domains.

\end{abstract}

\maketitle

\section{\label{sec:intro}Introduction}

The ability to detect weak magnetic fields with high sensitivity and spatial resolution forms an indispensable task across many areas of modern science and has the potential to enable transformative applications in fields ranging from fundamental physics\,\cite{budker2014,demille2017,bass2024quantum} and materials science \,\cite{ripka2021magnetic} to biological research \,\cite{mostufa2023flexible}. Over the past few decades, a variety of magnetic field sensors have been developed, some of which achieve remarkable levels of sensitivity \,\cite{shah2013compact,degen2017,mitchell2020colloquium}. Notable examples include superconducting quantum interference devices (SQUIDs)\,\cite{weinstock1996} and optically pumped magnetometers (OPMs)\,\cite{kominis2003subfemtotesla, budker2007, shah2013compact}, both of which can detect magnetic fields with sensitivities surpassing a femto-Tesla. However, these technologies typically suffer from significant limitations, such as the need for cryogenic operation, high power consumption, and limited spatial resolution.

As an alternative, individual nitrogen-vacancy (NV) centers in diamond have emerged as highly promising magnetic field sensors, offering nanoscale spatial resolution and operation under ambient conditions\,\cite{balasu2008,degen2008,rovny2024nanoscale}. These solid-state quantum sensors exploit the long coherence times ($\text{T}_2$) of NV center spins for the detection of oscillating (AC) magnetic fields\,\cite{taylor2008,balasu2009}, achieving sensitivities ($\eta_\text{ac}$) in the nano-Tesla range\,\cite{maze2008,balasu2009}. However, the sensitivity of NV centers to static (DC) magnetic fields is significantly lower—typically around 10\,$\mu$T—and is fundamentally limited by the shorter dephasing time ($\text{T}_2^*$, where $\text{T}_2^*<\text{T}_2$), such that $\eta_\text{dc} \propto 1/\sqrt{\text{T}_2^*}$ \,\cite{Barry-RMP-2020-NVmagnetometry}. This challenge is particularly pronounced for shallow NV centers embedded in diamond crystals with natural $^\text{13}$C abundance, which are commonly used for nanoscale sensing applications \,\cite{schirhagl2014nitrogen,casola2018probing}. In these cases, the proximity to surface-related noise sources further degrades $\text{T}_2^*$, thereby limiting their full potential for static magnetic field detection.

Hybrid sensing strategies, in which NV centers are integrated with complementary materials or sensing platforms to overcome the limitations of standalone NV systems and thereby achieve enhanced sensing performance, have attracted considerable attention in recent years. For example, NV centers have been integrated with piezoelectric materials to enhance magnetic sensitivity\,\cite{cai2014hybrid}, with magnetic materials to achieve high bandwidth\,\cite{carmiggelt2023broadband} and frequency conversion\,\cite{wu2024wideband}, and with vapor magnetometers to combine high sensitivity with vector magnetometry\cite{shalev2025quantum}.

 In this work, we present a hybrid approach that leverages the significantly longer $\text{T}_2$ time of shallow NV centers to achieve a $\eta_\text{dc}$ of 63\,nT/$\sqrt{\text{Hz}}$. This represents a substantial improvement over the conventional Ramsey method, which typically yields $\eta_\text{dc}$ on the order of tens of $\mu$T. Our method enhances $\eta_\text{dc}$ by exploiting near-field magnetic interactions between the NV center and a magnetic microwire exhibiting `giant magneto-impedance' (GMI) response. Notably, the demonstrated sensitivity of 63\,nT/$\sqrt{\text{Hz}}$ for a shallow NV center exceeds that of the best-reported single NV sensitivities in bulk diamond (approximately 300\,nT \,\cite{dreau2011avoiding}) despite the latter’s superior isolation from surface noise. This improvement highlights the potential of our approach for highly sensitive external magnetic field sensing with shallow NV centers.

\section{\label{sec:sensor descript}Hybrid sensor description}

Magneto-impedance response of certain soft ferromagnetic materials offers a highly sensitive mechanism for magnetic field detection\,\cite{phan2008giant}. In that context, Giant magneto-impedance or simply the GMI effect in ferromagnetic alloys is a classical phenomenon which occurs when a small change in the DC magnetic field ($H_{dc}$) in their vicinity causes a large changes in their AC complex impedance ($Z$)\,\cite{mohri1993a,kawashima93,panina1994,peng2016}. The GMI effect has its origin in the $\textit{skin effect}$ wherein the penetration depth of AC current depends on its frequency ($f_\text{ac}$) and the relative permeability  ($\mu_r$) of the material (see Appendix-\ref{sec:GMI concept} for more details).
In the so-called intermediate frequency range—from approximately $f_\text{ac} \approx$ 100\,kHz to $\approx $1\,GHz—materials exhibit electrical impedance that is highly sensitive to external DC magnetic fields \,\cite{knobel2002}. Within this range, even a small DC field on the order of a fraction of a milli-Tesla can produce a substantial change in impedance. The relative change is quantified by the GMI ratio:

\begin{equation}
\centering
\frac{\Delta Z}{Z} = \frac{Z(H_\text{dc}) - Z(H_\text{dc}^{'})}{Z(H_\text{dc}^{'})} \times 100\%
\end{equation}

where $\Delta Z$ is the change in impedance, and $Z(H_\text{dc})$ is the impedance at a given external DC magnetic field $H_\text{dc}$, and $Z(H'_\text{dc})$ is the impedance at a maximum field $H'_\text{dc}$, typically chosen to be large enough to saturate the magneto-impedance response. This GMI ratio can reach values as high as 500$\%$, highlighting the material’s strong magneto-impedance response even under weak magnetic fields\,\cite{phan2008giant}. A typical response of the GMI ratio $\left(\frac{\Delta Z}{Z}\right)$ as a function of external DC magnetic field amplitude is shown in figure-\ref{fig_s2}(a) (Appendix). This characteristic behavior is well established in the GMI research community. The pronounced field-dependent impedance response has enabled the development of highly sensitive magnetic field sensors capable of detecting fields down to pico-Tesla scale under ambient conditions \,\cite{peng2016,uchiyama2012,mohri2015}. However, the reliance on traditional induction-based readout schemes results in relatively bulky sensor designs, thereby limiting spatial resolution to the millimeter scale \,\cite{chen2018,yu2011,mohri2015,peng2016}. This size constraint presents a major challenge for the continued miniaturization of GMI-based sensor platforms.

On the other hand, nitrogen-vacancy (NV) centers in diamond are well established as highly sensitive quantum sensors for magnetic fields \,\cite{schirhagl2014nitrogen,rondin2014}. Due to their atomic-scale size, these sensors can be positioned within nanometers of a target, enabling spatial resolution at the nanometer scale\,\cite{balasu2008,taylor2008, maletinsky2012}.

In this work, we demonstrate a hybrid magnetic field sensing system that integrates a GMI wire, a classical sensor, with a quantum NV center sensor, achieving both high sensitivity and nanoscale spatial resolution. In the hybrid sensing scheme presented here, the NV center ground state couples to the RF magnetic field $B_{\text{rf}}(Z, t)$ generated by the GMI wire, whose amplitude is modulated by changes in the wire impedance $Z$ through the GMI effect. The corresponding NV ground-state Hamiltonian can be written as,

\begin{equation}
\hat{\mathcal{H}_{\text{NV}}} =
D\hat{S}_{z}^{2}
+
A_{\parallel}\hat{S}_{z}\hat{I}_{z}
+
A_{\perp}\left(
\hat{S}_{x}\hat{I}_{x}
+
\hat{S}_{y}\hat{I}_{y}
\right)
+
\gamma_{e}B_{0}\hat{S}_{z}
+
\gamma_{e}B_{\mathrm{rf}}(Z, t)\hat{S}_{z},
\label{eq:NVHamiltonian}
\end{equation}

where, $D$=2.87\,GHz is NV zero-field splitting parameter, $A_{\parallel}$ and $A_{\perp}$ are the longitudinal and transverse hyperfine coupling constants of the host $^{15}$N nucleus, $B_0$ is the static bias magnetic field ($\sim$0.5\,mT) produced by the GMI wire and aligned along the NV symmetry axis, $\gamma_e = 2\pi\times \text{28}\,\text{GHz}\cdot\text{T}^{-1}$ is the electron gyromagnetic ratio, and $\hat{S}$ and $\hat{I}$ are electron and nuclear spin operators, respectively. The dependence of $B_{\text{rf}}$ on $Z$ explicitly captures the GMI transduction mechanism underlying the hybrid sensor. Specifically, an external DC magnetic field to be detected modifies the wire impedance $Z$, which changes the amplitude of the RF magnetic field $B_{\text{rf}}$ generated by the GMI wire and subsequently detected by the NV center.

\section{Experimental Setup}

\begin{figure}
    \centering
    \includegraphics[width=0.99\linewidth]{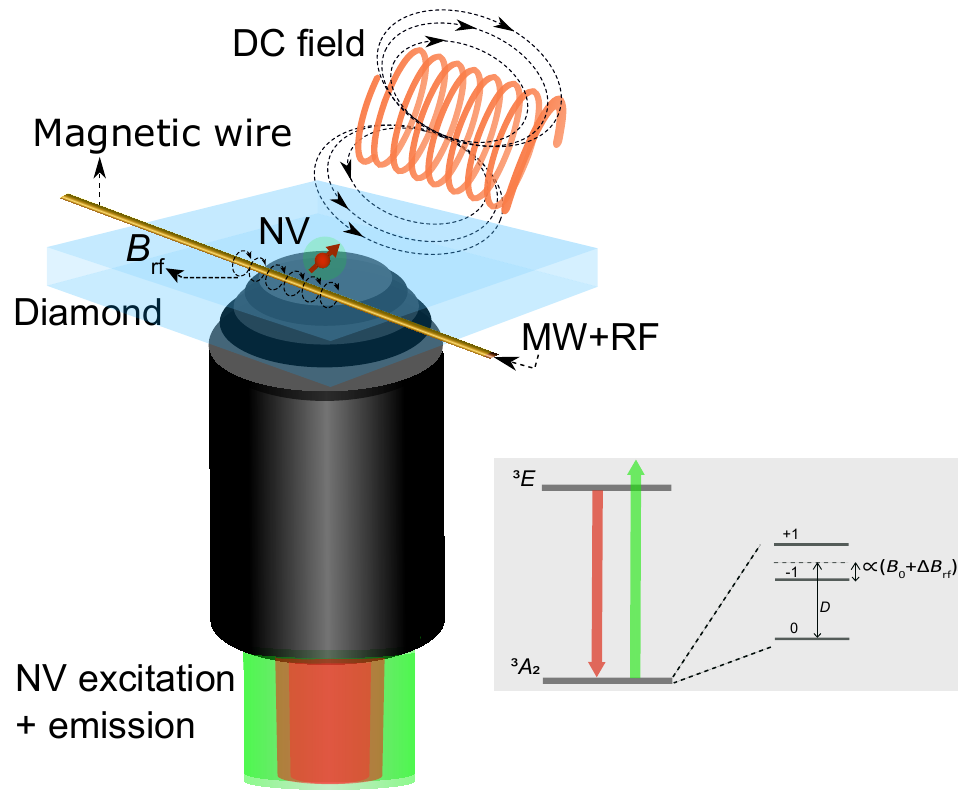}
    \caption{\small A simplified schematic of the GMI–NV-based hybrid sensor setup is shown. A soft ferromagnetic (GMI) wire is positioned on top of a diamond wafer containing individual shallow NV centers (indicated by a red sphere with an arrow). A microscope objective, aligned through the wafer, is used to focus a green excitation laser and collect the resulting NV fluorescence. Microwave pulses, used to manipulate the NV spin states, and a radio-frequency waveform, used to activate the GMI effect, are applied simultaneously through the GMI wire. A nearby coil generates the DC test magnetic field to be detected. Inset shows the simplified energy level diagram of NV, with its ground state coupled to magnetic fields ($B_0$, $B_{\text{rf}}$) from the GMI wire.}
    \label{fig1}
\end{figure}

To realize this, a GMI microwire was placed on top of a diamond chip containing shallow-implanted NV centers located a few nanometers beneath the surface \,\cite{kavatamane2019probing} (figure-\ref{fig1}(a)). We characterized the sensitivity of this hybrid NV-GMI sensor to external DC magnetic fields ($B_\text{dc}$).
A high numerical aperture (NA = 1.35) oil-immersion microscope objective was used to image individual NV centers, and confocal fluorescence images of the diamond sample were obtained. See Appendix-\ref{sec:exp setup} and -\ref{sec:diam sample} for details of the experimental setup and sample preparation. As illustrated in figure-\ref{fig1}(a), microwave (MW) pulses for NV spin manipulation and radio frequency (RF) signals to induce the GMI effect were simultaneously delivered to the wire via two separate channels of an arbitrary waveform generator. The GMI wire also serves as a microwave antenna, eliminating the need for a separate dedicated antenna (such as a copper wire) and thereby simplifying the experimental design.

A nearby coil, driven by a variable DC current source, generates the external $B_\text{dc}$ to be measured. The coil was placed at a distance of $\sim$10\,mm from the sample, with its axis aligned parallel to the longitudinal axis of the magnetic wire. In this configuration, the GMI wire responds to changes in the external magnetic field, and the response is read out with high sensitivity by a single NV quantum sensor.

\section{Results}

\subsection{Characterizing coherence of NV spin}

In this study we use a GMI wire of the composition Co\textsubscript{69.25}Fe\textsubscript{4.25}Si\textsubscript{13}B\textsubscript{12.5}Nb\textsubscript{1}. Since the wire is magnetic, there always exists a stray field in the vicinity of the wire\,\cite{eggers2017tailoring} (typical coercive field is $\sim 2$\,Oe). As a result, a NV center situated near the wire typically experiences a static field ($B_0$) of a few hundred $\mu$-Tesla originating from the magnetic domains of the wire. This static field acts as a small bias field which will split the degenerate $m_s=\pm1$ spin sublevels of the NV spin in its ground state. The presence of small stray static field eliminates the need for applying an external bias magnetic field for the NV center, thereby further simplifying the sensor setup. 

\begin{figure}
     \centering
        \includegraphics[width=\linewidth]{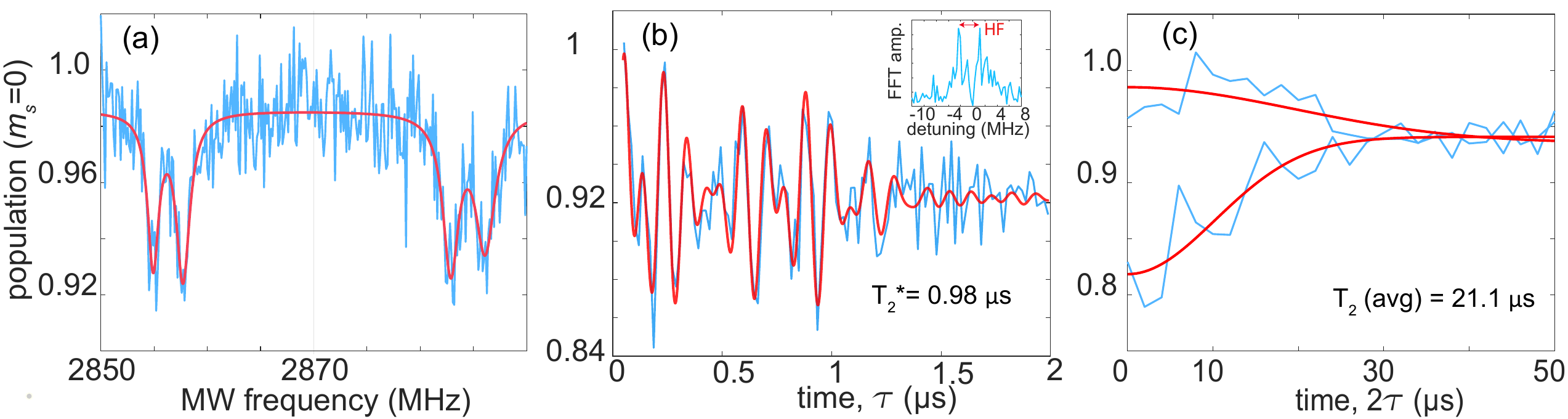}
\caption
{\small Electron spin resonance measurements of an individual shallow NV center in the presence of stray magnetic field from the magnetic wire.  (a) ODMR spectrum. Due to the presence of stray magnetic field from the wire ($B_0\sim$ 0.5\,mT), NV center's otherwise degenerate $m_s=\pm1$  spin sub levels are split from the zero field line $D$ =  2870\,MHz. Red solid line is a Lorentzian fit to the data. The two dips at a separation of $\sim$3\,MHz on either side of $D$ indicate the presence unpolarised $^{15}$N nuclear spin of the NV center. (b) Ramsey interference signal exhibiting a decay with a characteristic timescale of $\text{T}_2^*$ = 0.98\,$\mu\text{s}$. The red line represents a fit to $\exp[-(\tau/\text{T}_2^*)^2]\sum_{i=1}^{4}\cos(2\pi f_i \tau)$, where $f_i$ are the frequency shifts due to hyperfine couplings of the NV center to the host 
$^{15}$N and a nearby weakly coupled $^{13}$C nuclear spins. Inset: Fast Fourier transform (FFT) of the Ramsey signal as a function of detuning frequency, showing a hyperfine splitting (3\,MHz) corresponding to $^{15}$N nuclear spin, as identified in (a). (c) Hahn-echo measurement. The top and bottom blue traces correspond to echo signals with the phase of the final $\frac{\pi}{2}$ pulse shifted by 180$^\circ$ between the two traces. The red lines are fits to extract coherence decay. The average coherence time $\text{T}_2$ from both fits is 21.1\,$\mu\text{s}$.}

\label{fig2}    
\end{figure}

 Optically detected magnetic resonance (ODMR) is a straight forward way to measure the magnitude of the stray field at each NV center\,\cite{gruber1997}. By individually performing ODMR measurements on multiple NV centers located at various positions around the wire, a wide-field magnetic image of the wire can be reconstructed (see Appendix \ref{sec:MO wide field}). It is important to note that the static magnetic field generated by the wire exhibits a relatively complex spatial profile; therefore, the direction and magnitude of the field experienced by each NV center vary with position. However, this spatial inhomogeneity could be minimized by carefully tuning the GMI material properties and its composition. For the magnetometry experiments, we selected only those NV centers that exhibited symmetric and sufficiently large splitting in their ODMR spectra relative to the zero-field splitting parameter. Figure-\ref{fig2}(a) shows a representative ODMR spectrum for one such NV center. Unless otherwise stated, all data and analyses presented in this work correspond to this specific NV center.

 For all subsequent measurements, including magnetometry, we employ the NV spin transition between the $m_s = 0$ and $m_s = -1$ manifold. We obtained Rabi frequencies ($\Omega_{\text{R}}$) in the range of 10\,MHz. For measuring the spin dephasing timescale ($\text{T}_2^*$), a standard Ramsey measurement protocol was used, employing the pulse sequence $\frac{\pi}{2} - \tau - \frac{\pi}{2}$, inserted between two laser pulses--one for spin initialization and another for readout -- where $\tau$ is the variable free precession time. This measurement yielded a $\text{T}_2^*$ coherence time of 0.98\,$\mu$s (figure-\ref{fig2}(b)), which is typical for shallow NV centers prepared with standard diamond surface treatments (see Appendix-\ref{sec:diam sample}).

The Fourier transform of the Ramsey signal reveals hyperfine (HF) splitting (inset of figure-\ref{fig2}(b)), corresponding to the implanted $^{15}$N nuclear spin intrinsic to the NV center. A Lorentzian fit to the spectrum yields a full width at half maximum (FWHM) linewidth ($\delta\nu$) of 0.9 $\pm$ 0.12\,MHz for one of the resolved HF transitions. Once $\delta\nu$ is known, the minimum resolvable external magnetic field $\delta B$ can be estimated using the energy equation $h\delta\nu = g\mu_\text{B}\delta B$, where $h$ is Planck's constant, $g$ is the electron g-factor ($g = 2.0$), and $\mu_\text{B}$ is the Bohr magneton. Based on this equation, our measured linewidth would yield a photon shot-noise-limited magnetic field sensitivity of approximately 32\,$\mu$T.

Application of a Hahn-echo sequence  $\displaystyle \frac{\pi}{2} - \tau - \pi - \tau - \frac{\pi}{2}$ can effectively extend the coherence time from $\text{T}_2^*$ to a long-lived  $\text{T}_{2}$ timescale\,\cite{balasu2009,taylor2008}. In figure-\ref{fig2}(c) the data for Hahn-echo sequence was fitted to a stretched exponential function, $Ae^{-\left(\frac{\tau}{\text{T}_\text{2}}\right)^p}$, where $A$ is the signal amplitude and $p=2$, to extract the time constant $\text{T}_{2}$ value.  The data were acquired using final $\frac{\pi}{2}$ pulses with phases set to 0$^\circ$ and 180$^\circ$. As shown in the measurements, the two datasets exhibit different decay rates. Fitting the data yields $\mathrm{T}_{2}$ values of 14.5\,$\mu$s and 27.7\,$\mu$s for the up and down traces, respectively, corresponding to an average value of 21.1\,$\mu$s. Such discrepancies in the decay behavior are commonly attributed to imperfections in the microwave drive pulses, which may also account for the observed differences in this case. Notably, a $\mathrm{T}_{2}$ value of 21.1\,$\mu$s represents an enhancement of the coherence time by $\approx$ 30 compared to the $\mathrm{T}_2^*$ value measured in figure-\ref{fig2}(b). As described below, our sensing method exploits this relatively long lived coherence time ($\text{T}_{2}$) for sensing static external field. We further note that the reduced ODMR contrast in figure-\ref{fig2}(a), compared to the echo contrast (figure-\ref{fig2}(c), is attributed to a non-optimal microwave $\pi$-pulse duration ($\mathrm{T}_\pi > \mathrm{T}_2^*$) used in the ODMR measurement.

\subsection{Demonstrating high sensitivity static field sensor}

\begin{figure}
   \centering
      \includegraphics[width=\linewidth]{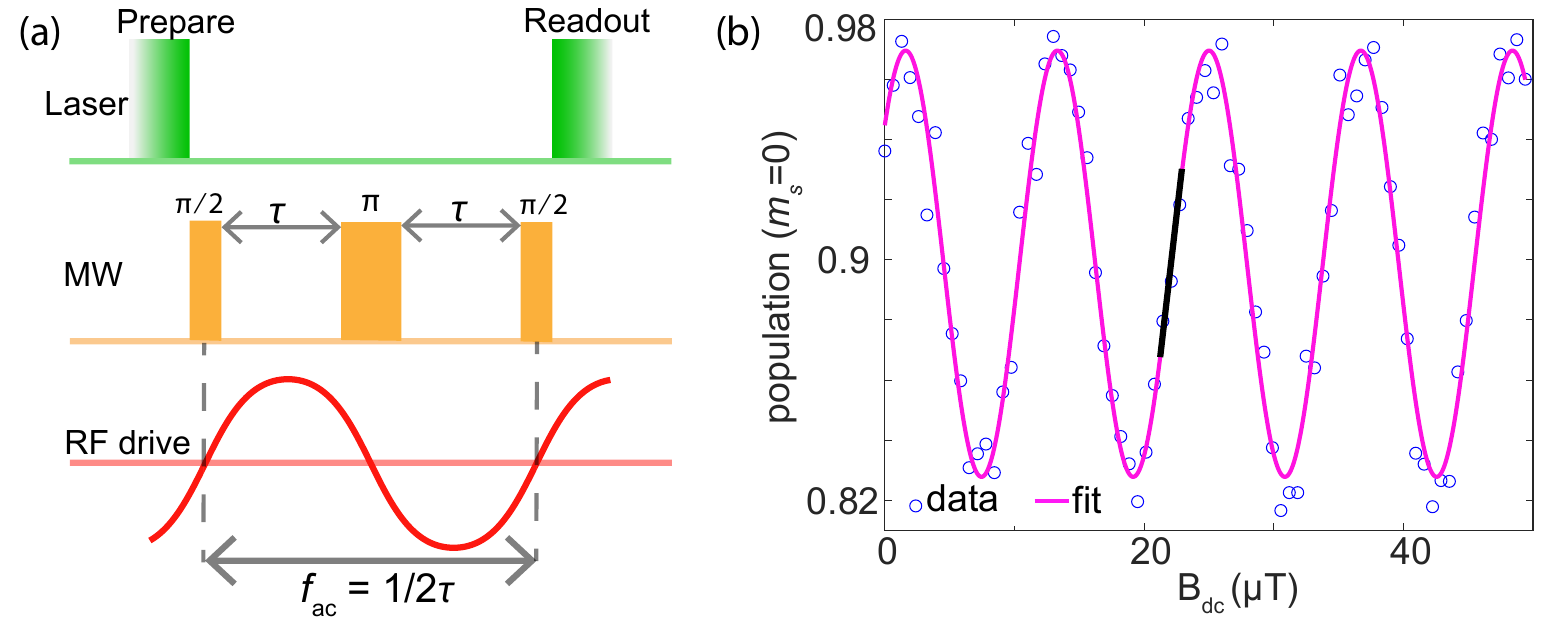}
\caption
{\small Demonstration of DC magnetic field sensitivity of the NV-GMI hybrid sensor.  (a) Schematic representation of the DC magnetometry. A Hahn-echo sequence with a fixed $\tau$ was applied to the NV. The GMI wire was simultaneously driven by a RF signal whose frequency was matched to the echo period ($f_\text{ac}=1/2\tau$)  (b) Oscillatory response of the magnetometer signal as a function of the amplitude of the external DC magnetic field. The RF drive frequency, $f_\text{ac}$ was set to 100\,kHz while its amplitude was kept at 1.0\,V. The data was fitted to a sinusoidal function. The sensitivity of the sensor is calculated from the maximum slope (black straight line) of the response. We obtain a sensitivity of 63\,$\mathrm{nT}/\sqrt{\mathrm{Hz}}$.}
\label{fig3}
\end{figure}

\normalsize

To detect weak external magnetic fields, we employ a Hahn-echo-based magnetometry technique integrated with a hybrid NV-GMI sensor. In conventional Hahn-echo sequences, the accumulated quantum phase of the NV center over the free precession periods cancels out, rendering the method inherently insensitive to static magnetic fields \,\cite{maze2008,balasu2009}. As a result, the NV spin ideally returns to its original state ($m_s=0$) at the end of the sequence.

In our approach, however, we circumvent this insensitivity by leveraging the field-dependent response of the GMI wire\,\cite{eggers2017tailoring}. The GMI wire acts as a transducer, converting variations in external static fields into changes in its impedance, which in turn modulates the local magnetic environment sensed by the NV center during the echo sequence.

The experimental protocol involved a standard Hahn-echo sequence with a fixed precession interval $2\tau$ (see figure-\ref{fig3}(a)), applied to the NV center. Concurrently, a time-varying RF drive voltage, $v(t') = v_{\text{ac}} \sin(2\pi f_{\text{ac}} t' + \phi')$, was applied to the GMI wire, where $v_{\text{ac}}$, $f_{\text{ac}}$, and $\phi'$ denote the amplitude, frequency, and phase, respectively. The RF frequency was chosen to satisfy $f_{\text{ac}} = 1/2\tau$, ensuring synchronization with the echo sequence. Importantly, the value of $f_{\text{ac}}$ was selected such that it falls within the optimal operating regime of the GMI effect—i.e., where the $Z$ of the wire exhibits strong sensitivity to external DC magnetic fields\,\cite{mohri1993a,kawashima93,panina1994}. The phase $\phi'$ was fixed at zero for all measurements.

Under this RF excitation, the GMI wire exhibits enhanced sensitivity to nearby static magnetic fields. Changes in the local DC magnetic field alter the wire’s $Z$, thereby modulating its magnetic behavior (i.e., $B_\text{rf}$). Since the NV center couples magnetically to the GMI wire, these $Z$-induced changes influence the NV’s spin evolution during the echo sequence, ultimately altering its fluorescence signal. We consider the linear operating regime of the GMI sensor, in which small variations in the external DC magnetic field, $\Delta B_\text{dc}$, produce proportional changes in the RF magnetic field generated by the GMI wire, i.e., $\Delta B_\text{rf}\propto\Delta B_\text{dc}$. Under a Hahn-echo sequence of fixed evolution time $2\tau$, the NV spin accumulates a phase $\delta\varphi$ determined by the instantaneous amplitude of $B_\text{rf}$. Consequently, varying the external $B_\text{dc}$ modulates the accumulated phase through the corresponding change in $B_\text{rf}$, giving rise to an oscillatory optical readout signal. The probability of finding the NV center in the $m_s=0$ state is therefore given by\cite{maze2008, balasu2009}, $P_0(B_\text{rf}) = \frac{1 + C(2\tau)\cos(\delta\varphi)}{2}$, where $\delta\varphi = \gamma_e\cdot B_\text{rf}\cdot2\tau$, $C(2\tau)$ is the Hahn-echo contrast in the absence of an external magnetic field, and $2\tau\leq\text{T}_2$ for the Hahn-echo sequence employed here.

To experimentally demonstrate the magnetometry, we gradually varied the amplitude of an externally applied static magnetic field $B_{\text{dc}}$ from 0 to 50\,$\mu$T and observed periodic oscillations in NV fluorescence intensity (figure-\ref{fig3}(b)). Note that since the probed magnetic field range is smaller ($\sim$ 0-50\,$\mu$T), the GMI response can be considered linear for this specific purpose. However, the GMI response of this material is non-linear for larger external magnetic field range as illustrated in figure-\ref{fig_s2}(a) of Appendix.  This fluorescence modulation encodes information about the field strength and enables quantitative field sensitivity analysis. The minimum detectable magnetic field $\text{B}_{\text{min}}$ was extracted using:

\begin{equation}
 \text{B}_{\text{min}} = \frac{\sigma_s}{\left( \frac{\text{d}S}{\text{d}B_{\text{dc}}} \right)},   
 \label{eq2}
\end{equation}

where $\sigma_s$ is the standard deviation of the fluorescence signal ($S$), and $\frac{\text{d}S}{\text{d}B_{\text{dc}}}$ is the maximum slope of the signal response curve shown in figure-\ref{fig3}(b). For an RF driving frequency of 100\,kHz (corresponding to $2\tau = 10 \, \mu\text{s}$), we achieved a static field sensitivity of $\eta_{\text{dc}} = \text{B}_{\text{min}} \sqrt{t} = 63\, \text{nT}/\sqrt{\text{Hz}}$, where $t$ is the total measurement time (see Appendix-\ref{sec:sens analysis} for error analysis). This sensitivity represents a 500X improvement over the estimated Ramsey-based DC field sensitivity (32\,$\mu$T) mentioned earlier for the same NV center, demonstrating the enhanced performance of the hybrid NV-GMI magnetometer in detecting weak static magnetic fields. This enhanced $\eta_{\text{dc}}$ of our sensor arises from two key factors. First, the GMI wire intrinsically exhibits a high sensitivity to external static magnetic fields far exceeding the Ramsey-based NV sensitivity. Second, the use of an echo sequence, rather than a Ramsey sequence, significantly extends the NV coherence time by mitigating inhomogeneous dephasing. This extension beyond the $\text{T}_2^*$-limit allows for longer interrogation times and thus improves the overall $\eta_{\text{dc}}$.

\begin{figure}
   \centering
      \includegraphics[width=\linewidth]{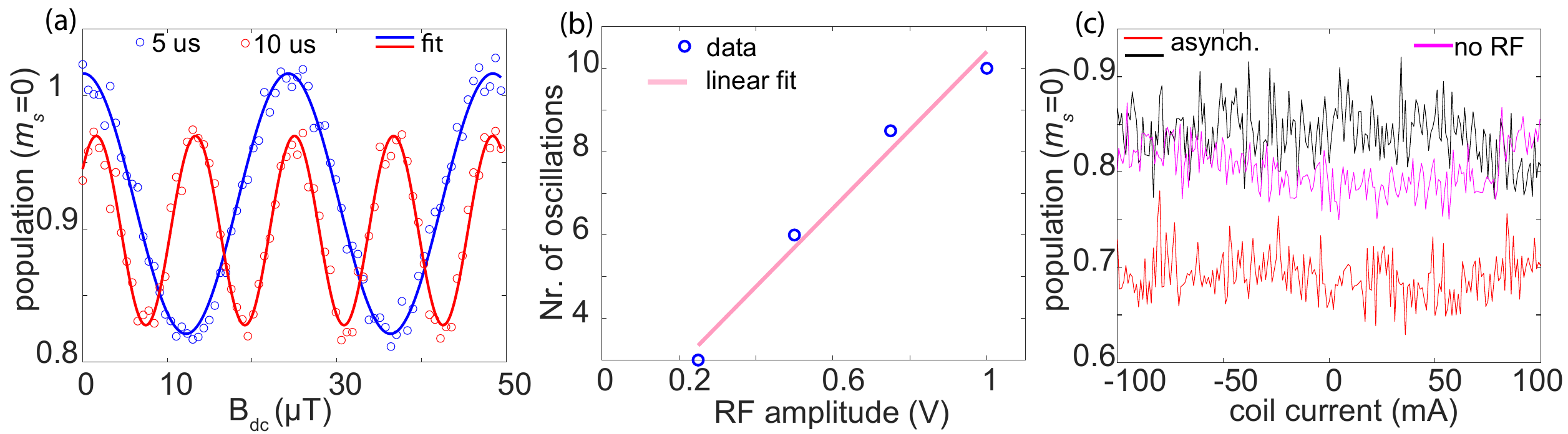}
\caption
{\small(a) NV-GMI hybrid magnetometer signal for different RF drive frequency (equivalently $1/\text{2}\tau$). Decreasing the $\tau$ gives higher signal contrast but at the expense of halved phase acquired by the NV and hence the reduced $\eta_{\text{dc}}$. Applying the same sensitivity analysis described in the main text [Eq.\,(\ref{eq2})], we obtain a DC magnetic field sensitivity of $\eta_{\mathrm{dc}} \approx \text{130}\,\mathrm{nT}/\sqrt{\mathrm{Hz}}$ for the case of 5\,$\mu\mathrm{s}$ signal.  (b) Number of oscillations of the magnetometer signal vs. amplitude of the RF drive voltage across the GMI wire. (c) Magnetometer signal as a function of external $B_\text{dc}$ amplitude for three different conditions: no RF drive applied to the GMI wire (magenta), RF drive with asynchronous frequency (black), and RF drive with asynchronous phase (red). For clarity, the traces are vertically offset. Note that $B_\text{dc}$ values are represented in terms of the raw current applied to the field-generating coil. For (a) and (c), the RF amplitude was maintained at a maximum value of 1.0\,V.}
\label{fig4}
\end{figure}

\normalsize
Note that we chose a nominal value of 2$\tau$ (1/$f_\text{ac}$= 10\,$\mu$s) where the NV center's coherence is sufficient to provide a good signal-to-noise ratio in the spin contrast and which also allows for the GMI effects to occur. If we increase the $\tau$, number of oscillations in the magnetometer signal correspondingly increases at the cost of reduced NV spin contrast (figure-\ref{fig4}(a)) due to loss of coherence. We note that the echo data shown in figure-\ref{fig3}(b) and \ref{fig4}(a) exhibit reduced noise compared to the echo decay signal in figure-\ref{fig2}(c), primarily due to the larger number of experimental repetitions performed in those measurements, resulting in enhanced signal-to-noise ratio. Since the number of oscillations in the magnetometer signal directly contributes to improved sensitivity, maximizing these oscillations becomes desirable. The number of oscillations scales linearly (figure-\ref{fig4}(b) with the RF drive amplitude. Therefore, throughout all measurements, the sensor was operated at the maximum allowable RF amplitude to optimize sensitivity. 

To confirm that the observed oscillations in the magnetometer signal originate from the NV center's magnetic interaction with the GMI wire, and not from experimental artifacts, we performed control experiments with varying conditions. First, we set the RF drive amplitude to zero and applied a fixed-$\tau$ echo sequence (same scheme as in figure-\ref{fig3}(a)) to the NV center while varying the external $\text{B}_{\text{dc}}$ amplitude in discrete steps. At each step, the corresponding NV fluorescence was recorded. In the absence of RF excitation, the fluorescence signal remained flat, showing almost no oscillatory behavior (figure-\ref{fig4}(c)). In contrast, when the RF drive was turned on, oscillations were revived, and their number increased with increasing RF drive strength (figure-\ref{fig4}(b)). Furthermore, no oscillations were observed when the RF signal was applied asynchronously with respect to the NV echo sequence--either in frequency ($f_\text{ac}\neq 1/2\tau$) or in phase ($\phi'\neq 0$) as seen in figure-\ref{fig4}(c). This indicates a coherent interaction exists between the GMI element and NV, and a precise synchronization of the RF drive with the NV echo sequence is needed. These results confirm that the oscillations arise from the near-field magnetic coupling between the NV center and the GMI wire, mediated by the RF-induced modulation of the wire’s $Z$.

\subsection{Noise floor}

\begin{figure}
   \centering
      \includegraphics[width=0.5\linewidth]{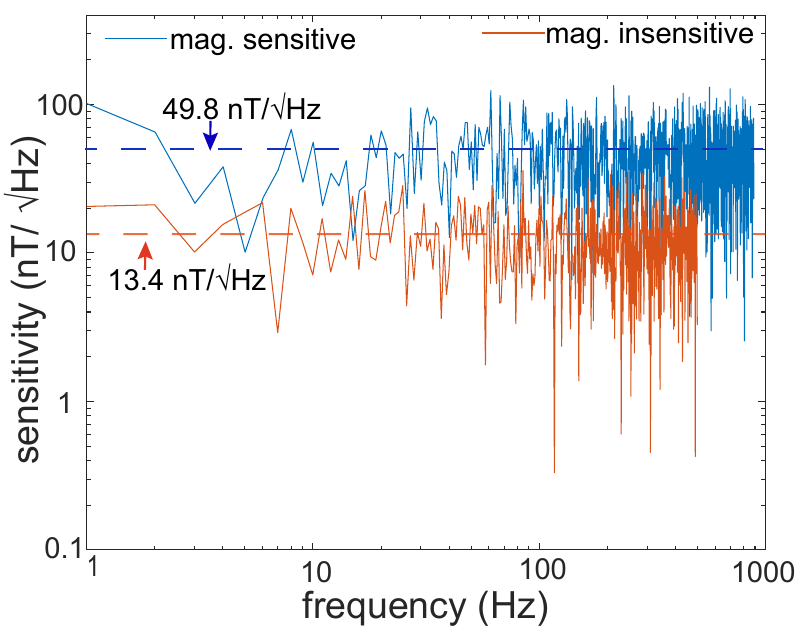}
\caption
{\small Noise spectral density of the hybrid NV-GMI magnetometer. The sensor was operated at maximum sensitivity by applying the RF drive signal at optimal frequency and maximum amplitude, with the Hahn-echo sequence synchronized to the drive. The magnetically sensitive trace was recorded under the application of a small static test field.} 
\label{fig5}
\end{figure}

Figure-\ref{fig5} presents the noise spectral density of the hybrid NV-GMI magnetometer. For the magnetically sensitive configuration, the sensitivity exhibits a mean noise floor of $\approx$ 49.8\,nT/$\sqrt{\text{Hz}}$. These data were acquired using the magnetometry protocol described in figure-\ref{fig3}(a), with a constant but small value of external $\text{B}_{\text{dc}}$ applied via the field-generating coil (see Appendix-\ref{sec:sens analysis}). To our knowledge, the best reported DC field sensitivities for GMI-based sensors lie in the range of 1–100\,pT/$\sqrt{\text{Hz}}$ \,\cite{malatek2013improvement}. Therefore, there is still a room for improvement in sensitivity for the hybrid sensing modality presented in this work. Nevertheless, this modality offers a highly sensitive platform for DC magnetometry with potentially nanoscale spatial resolution through further miniaturization of the GMI sensing element—something not achievable with conventional GMI sensors alone. For comparison, figure-\ref{fig5} also shows the background noise spectrum with a mean noise floor value of $\approx$ 13.4\,nT/$\sqrt{\text{Hz}}$, recorded with all microwave, RF, and laser sources turned off. This background originates primarily from ambient light leakage into the photon detector and exhibits a slightly lower noise floor than the magnetically sensitive case.

\section{Discussion}

The NV-GMI-based Hahn-echo magnetometry demonstrated in this work serves as a proof-of-concept, and there is considerable scope for further improving the $\eta_{\text{dc}}$ of this hybrid sensing platform. One promising avenue is to extend the NV center's coherence time $\text{T}_\text{2}$ beyond what is achievable with the Hahn-echo sequence by employing dynamical decoupling protocols, which would directly enhance sensitivity. Additionally, implementing double-quantum magnetometry techniques \,\cite{mamin2014multipulse} could, in principle, offer up to a 2-fold improvement in $\eta_{\text{dc}}$. A more substantial enhancement, potentially an order of magnitude, can be gained by improving the NV photon collection efficiency. This can be achieved through nanophotonic engineering, such as the use of diamond nanopillar waveguides \,\cite{momenzadeh2015nanoengineered} or circular diamond gratings \,\cite{li2015efficient}. Further gains are possible by adopting spin-to-charge conversion techniques \,\cite{shields2015efficient} or by absorption of NV singlet states\,\cite{zadeh2024master}, which enable higher-fidelity spin readout than that is achieved with NV fluorescence detection. With these efforts, the projected sensitivity of the hybrid NV-GMI sensor could reach the level of a few nT/$\sqrt{\text{Hz}}$, approaching the AC field sensitivity of single NV centers \,\cite{balasu2009}—but notably, without the need for specially engineered or high-purity diamond samples.
Instead of single NV centers, if ensemble NV centers are used then the $\eta_{\text{dc}}$ of the
resulting hybrid sensor would scale as $1/\sqrt{N}$ ($N$ is the number of NVs in the sensing volume), at the expense of spatial resolution. It is worth noting that the $\eta_{\text{dc}}$ of the hybrid-sensor can further be tuned and optimized by varying the dimension and the GMI characteristics of the wire itself. Overall, given the state-of-the-art ensemble NV sensors offer $\eta_{\text{dc}}$ below a pico-Tesla\,\cite{barry2024sensitive}, a further improvement in $\eta_{\text{dc}}$ of a hybrid NV-GMI sensor could take it close to spin projection limit, thus being on-par with SQUIDs and OPMs. 

It should be noted that the GMI wire used in the present study was an off-the-shelf component and was not specifically engineered for implementation in a hybrid sensing platform. Consequently, the spatial magnetic field profile surrounding the wire exhibits a relatively complex structure (see Appendix~\ref{sec:MO wide field}) prompting us to choose an NV with highest splitting in the ODMR spectrum, which may at first appear to limit the applicability and scalability of this approach. However, this complexity primarily reflects the non-optimized nature of the material. Through targeted tuning of the GMI material properties and composition, it should be possible to achieve a more homogeneous magnetic field distribution around the wire, thereby enabling the selective and efficient use of NV centers with a desired crystallographic orientation. In addition, the diamond sample employed in this study contains NV centers aligned along all four crystallographic orientations. Utilizing diamond substrates with preferentially oriented NV centers would improve the scalability of the hybrid sensing architecture demonstrated here; for an ensemble based sensor, this preferential orientation also reduces spectral overlap of different NV orientations and simplifies the readout, thereby enhancing signal contrast.

A sensitive NV-GMI magnetometer is well-suited for detecting static or slowly varying magnetic fields in the following areas. In biomedical applications, it could enable high-resolution, extracellular detection of neuronal activity \,\cite{barry2016optical}; in medical diagnostics, perform magnetoencephalography and magnetocardiography under ambient conditions, which require $\sim100\,\text{fT}/\sqrt{\text{Hz}}$\,\cite{matti1993} and $\sim1\,\text{pT}/\sqrt{\text{Hz}}$\,\cite{kwong2013diagnostic}, respectively— thus offering a cryogen-free alternative to SQUID-based systems while substantially reducing the sensor-to-specimen stand-off distance. Furthermore, scaling down from the micrometer-scale GMI wire used in this study to a thin film or a nanowire, and integrating it with a single NV center, could pave the way for highly sensitive detection of magnetically functionalized biomolecules through their binding to the GMI element (see ref.\,\cite{chiriac2005magnetic})—potentially enabling single-molecule sensitivity. A compact and highly sensitive nature of the NV-GMI sensor would potentially be advantagesous for practical deployment of quantum sensors in applications such as geophysical surveying, mineral exploration, and precision navigation. 

\section{Conclusion}
In summary, we have presented a hybrid DC magnetic field sensor that leverages the high sensitivity of a magneto-impedance material and the quantum properties of near-surface NV centers in diamond under ambient conditions. This approach achieves field sensitivities over two orders of magnitude greater than conventional $\text{T}_2^*$-based NV magnetometers. By eliminating the need for a separate microwave antenna and bias field, the sensor significantly simplifies the experimental setup and enables a compact footprint. Looking ahead, further miniaturization of the GMI sensing element could enable nanoscale spatial resolution in this platform, making it a promising route toward advanced, deployable diamond-based quantum sensors for high-resolution detection of slowly varying magnetic fields in a broad range of applications.

\begin{acknowledgments}
The authors (VKK, DD, and GB) thank the Max-Planck-Gesellschaft and Niedersächsisches Ministerium für Wissenschaft und Kultur for funding and support. VKK thanks Paul Barclay for the financial support during the preparation of manuscript and Nicolas Sorensen for discussions. 
\end{acknowledgments}

\appendix

\section{\label{sec:exp setup}Experimental setup}

Experiments were carried out on a home-built confocal microscope using a 532\,nm green laser (gem532, Laser Quantum) for NV excitation and an oil-immersion objective (UPlanSApo 60X, NA:1.35) for focusing the laser onto the sample and collecting the NIR fluorescence from it. Following standard confocal method, the fluorescence was focused to a 25\,$\mu m$ pinhole and detected using two avalanche photo detectors (SPCM-AQRH, EXCELITAS) operating in the Hanbury Brown and Twiss anti-bunching configuration to determine the single NV centers. An acousto-optic modulator, AOM (AA opto-electronic, 200\,MHz center frequency) was used in a double-pass scheme to obtain the pulsed laser for the NV spin manipulation. This double-pass AOM pulsing typically achieved a laser extinction ratio (measured by the leaking of laser light when the AOM was turned off) to be $> 30\,\text{dB}$. A two-channel Arbitrary Waveform Generator, AWG (model:AWG70002A, Tektronix), operating at a sampling rate of 20\,Gsps is the main control unit of our experimental setup. It generates the optical pulses (via AOM),  microwave (MW) pulses and the radio frequency driving signals necessary for the experiments. The MW pulses generated in one channel of the AWG are amplified using a 16\,W high power amplifier (minicircuits) and applied to the `RF port' of a Bias-Tee Diplexer connector (200B-FF-3, MECA Electronics). The RF driving signal with a maximum amplitude (variable over 0-1\,V, peak-to-peak), for driving the GMI wire was generated at the second channel of the AWG and applied to `DC or low frequency port' of the Bias-Tee connector. The output port of the Bias-Tee was then connected to one end of the GMI wire. This way we were able to simultaneously deliver the MW and the RF signal to the GMI wire. The GMI wire was carefully positioned on top of the diamond wafer, with both ends soldered to a home-built microwave stripline circuit designed for 50\,$\Omega$ impedance matching. One end of the wire was connected to a 20\,W fixed attenuator (Mini-Circuits BW-S20W20+), followed by a 50\,$\Omega$ termination (Mini-Circuits ANNE-50+). This setup typically yielded NV center Rabi frequencies of approximately 10\,MHz, corresponding to a $\pi$-pulse duration of about 50\,ns.
The test magnetic field ($B_\text{dc}$) for the magnetometry experiments was produced by passing a known DC current ($I_{DC}$) from a programmable current source (Keithley Model 6221) through a coil positioned near the micro-wire.

\section{\label{sec:diam sample}Diamond sample}

\begin{figure}[t]
   \centering
    \includegraphics[width=0.75\linewidth]{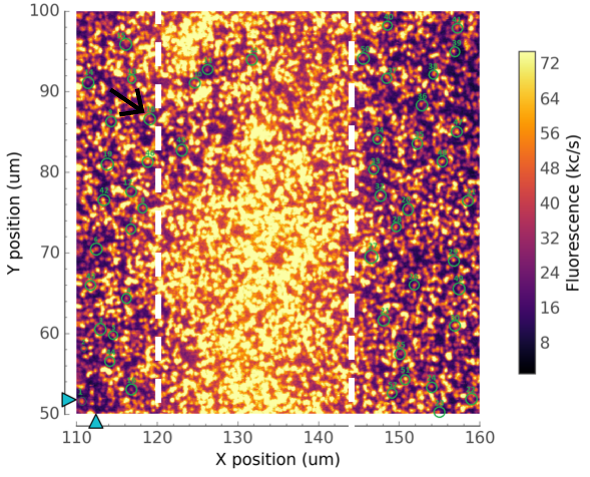}
    \caption{Confocal photoluminescence scan (50\,$\mu$m $\times$ 50\,$\mu$m) of the shallow implanted NVs around the magnetic wire (indicated by white dashed lines). ODMR results from over 70 NV centers were used to reconstruct the magnetic field profile from the wire. Some of the NV centers used for the measurement are shown in green circles. The NV$\#$18 (shown by black arrow near top left corner) was used for the results discussed in the main article.} 

\label{fig_s1}
\end{figure}

Shallow NV centers in the high purity (1.1$\%$ of $^{13}$C concentration) CVD diamond (from element six) of dimensions 3\,mm $\times$3\,mm $\times$0.1\,mm were produced by implantation of $^{15}$N ions\,\cite{kavatamane2019probing}. For implantation we used a commercial implanter (SPECS surface Nano Analysis GmbH). We used $^{15}$N ions at a dose of 10$^9$ ions/cm$^2$  and low energies  (2.0\,keV) for the shallow implantation. To form NV centers the sample was then subjected to high temperature annealing  (800$^{\circ}$C) under high vacuum conditions  (10$^{-7}$\,mbar) using a commercial vacuum setup  (minicoater, tectra GmbH) capable of controllably ramping up/down the temperature. This procedure yielded NV densities good enough to isolate the individual centers under the confocal fluorescence microscope. We expect the NV centers to be at a few nm below the diamond surface based on the standard SRIM simulations\,\cite{kavatamane2019probing}. Prior to the experiments, the sample was boiled in a mixture of three acids (sulfuric, nitric, and perchloric acid) at 1:1:1 ratio for several hours. In our case, this acid treatment produces spin dephasing times  ($\text{T}_2^*$) of single NV centers to be typically around 1\,$\mu$s and leaves the surface oxygen terminated. 

Figure-\ref{fig_s1} shows a confocal image of shallow implanted NV centers in an area of 50\,$\mu$m $\times$ 50\,$\mu$m around the GMI wire.

\section{\label{sec:GMI concept} Giant magneto-impedance (GMI) concept}

Giant magneto-impedance or GMI is a high frequency analogue of well known giant magnetoresistance (GMR), which is the large change in the electrical resistance of a material upon change in the external magnetic field. In both cases (GMI and GMR) one observes a large change in the voltage across the material when the magnetic field is applied. However, GMI and GMR are entirely different concepts in that GMI is purely a classical phenomenon whereas the GMR effect has quantum mechanical origins. The interest in GMI is spurred by several advantages over the conventional magnetic sensors, namely the high sensitivity ($<$\,nT) and room temperature operation.

\begin{figure}[t]
    \centering
    \includegraphics[width=\linewidth]{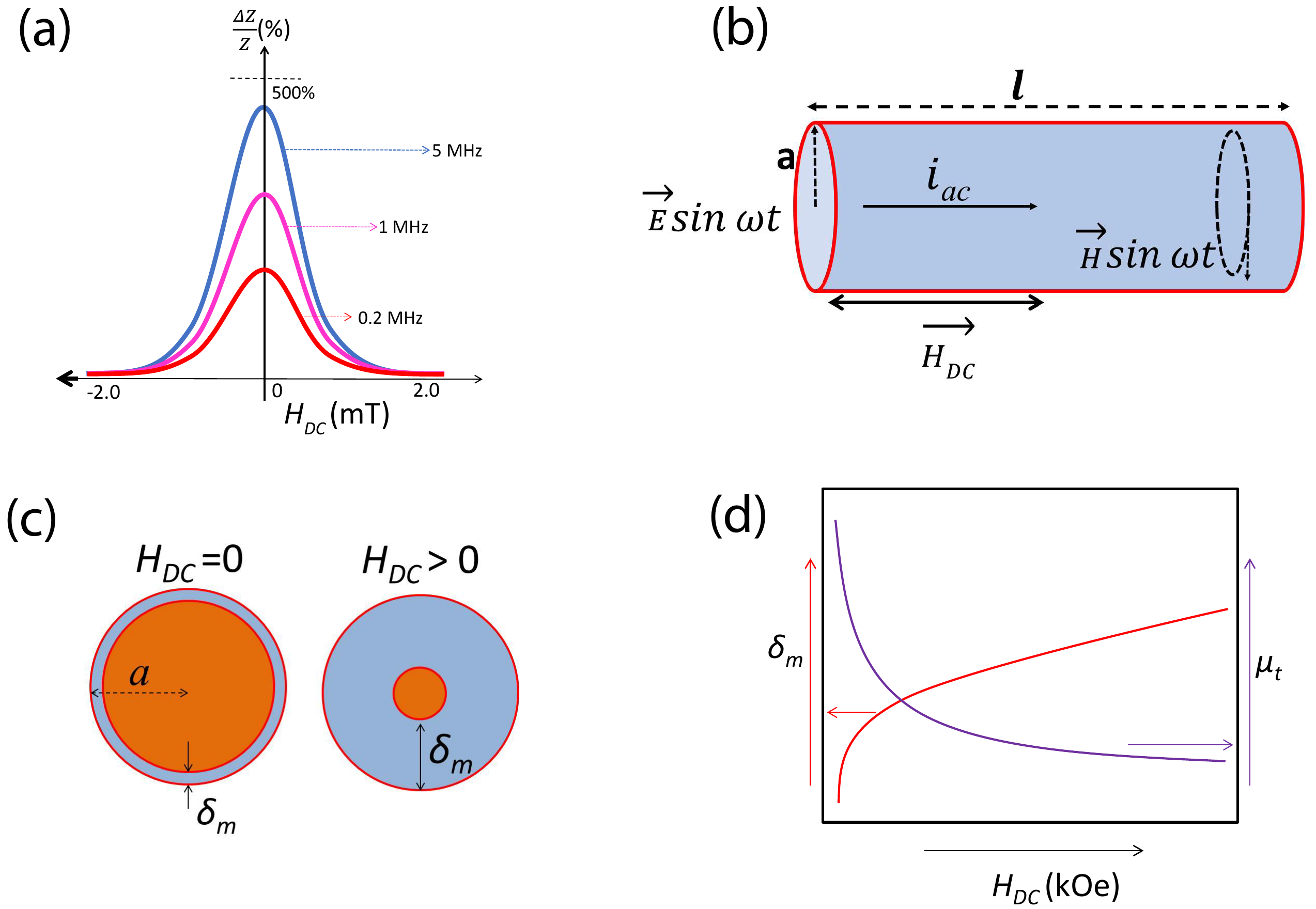}
    \caption{(a) An illustration of a typical frequency and DC magnetic field response of the GMI ratio $\frac{\Delta Z}{Z}$. Large change in the $\frac{\Delta Z}{Z}$ for small changes in the $H_{DC}$ and its dependence on frequency of radio frequency drive signal can be seen. (b) A schematic illustration of the GMI wire, with length $l$ carrying the alternating current $i_{ac}$ and subjected to static external test field, $H_{DC}$. (c) An illustration of variation skin depth $\delta_m$, with magnitude of $H_{DC}$, as seen from the cross section of the GMI wire with a radius of $a$. (b) Variation of skin depth (left) and total permeability ($\mu_t$) of the wire as a function of $H_{DC}$.}
    \label{fig_s2}
\end{figure}

Giant magnetoimpedance is defined as a phenomenon in which under the application of a small alternating current ($i_{ac}$), the AC complex impedance ($Z$) of a soft ferromagnetic conductor undergoes a large change when a tiny static external magnetic field ($H_{DC}$) is applied. This effect is graphically illustrated in figure-\ref{fig_s2}(a) while in figure-\ref{fig_s2}(b) a cylindrical GMI wire under the application of $i_{ac}$ and $H_{DC}$ is schematically shown. More details on the material characterization are given in reference\,\cite{eggers2017tailoring}.

The concept of magneto-impedance (MI) has its origin in the well known Skin effect in electromagnetism. The alternating radio frequency (RF) current $i_{ac} = i_0e^{jwt}$ in a conductive element such as a metal wire, tends to be distributed near the surface (skin) of the conductor (figure-\ref{fig_s2}(c)). This means the current density is not homogeneous over the cross section of the conductor but largest near the surface and decreases exponentially away from the surface. The skin depth ($\delta_m$), the region below the outer surface of the conductor where the effective current flows is given by,

\begin{equation}
    \delta_m = \sqrt{2\rho/\omega\mu}
    \label{eq1}
\end{equation}

where, $\rho$ is resistivity of the material, $\omega$, the angular frequency of the alternating current,
and $\mu_t$=$\mu_{r}\mu_{0}$ (where, $\mu_r$ and $\mu_0$ being the relative permeability and permeability of free space, respectively) is the total permeability. From equation-\ref{eq1} we can expect that as the $\omega$ increases, $\delta_m$ decreases which means that the $i_{ac}$ tends to be concentrated near the surface of the conductor. For ordinary non-ferromagnetic conductors such as copper, $\mu_t$ is independent of $\omega$ and the external magnetic field. But for ferromagnetic materials, $\mu_t$ strongly depends on $\omega$, amplitude ($i_0$) of the AC driving current as well as the direction and magnitude of the external applied magnetic field (figure-\ref{fig_s2}(d)). In soft ferromagnetic materials (materials which can be easily magnetized or demagnetized), the permeability $\mu_t$ changes by orders of magnitude even for the application of small external magnetic fields. The permeability $\mu_t$ is also sensitive to temperature, mechanical strain, and magnetic anisotropies in the material.

\subsection*{GMI material information}

For our studies we used a Co-rich amorphous GMI wire with a composition of \newline Co\textsubscript{69.25}Fe\textsubscript{4.25}Si\textsubscript{13}B\textsubscript{12.5}Nb\textsubscript{1} which was fabricated by melt-extraction method\,\cite{wang2011}. These wires are cylindrical in shape and known to exhibit soft ferromagnetic properties\,\cite{wang2017}. The GMI wire used in this work was fabricated in-house using the melt-extraction technique. Details of the fabrication procedure, material composition, and impedance characterization are provided in Refs.\,\cite{eggers2017tailoring,wang2011}.
The wires with a diameter of $\sim25\,\mathrm{\mu m}$ and length $\sim$30\,mm were used for the experiments. These wires coated with a thin layer (a few microns) of gold using standard electroplating method using a home-built setup involving gold plating solution and a current source. The end to end resistance of the wire was measured to be 10.2\,$\Omega$ after the soldering on the stripline. This means for a RF signal with a maximum applied amplitude of 1\,V corresponds to an AC current with amplitude of $\sim$100\,mA. We note that the shallow NV centers investigated in this work exhibited fluorescence rates and spin coherence properties ($\text{T}_2^*$ and $\text{T}_2$) after installation of the GMI wire that were comparable to those typically observed on a freshly cleaned diamond surface. This observation suggests that the presence of the GMI wire does not noticeably affect the NV spin properties.

\section{\label{sec:MO wide field}Magneto-optical wide field image}

\begin{figure}
    \centering
    \includegraphics[width=\linewidth]{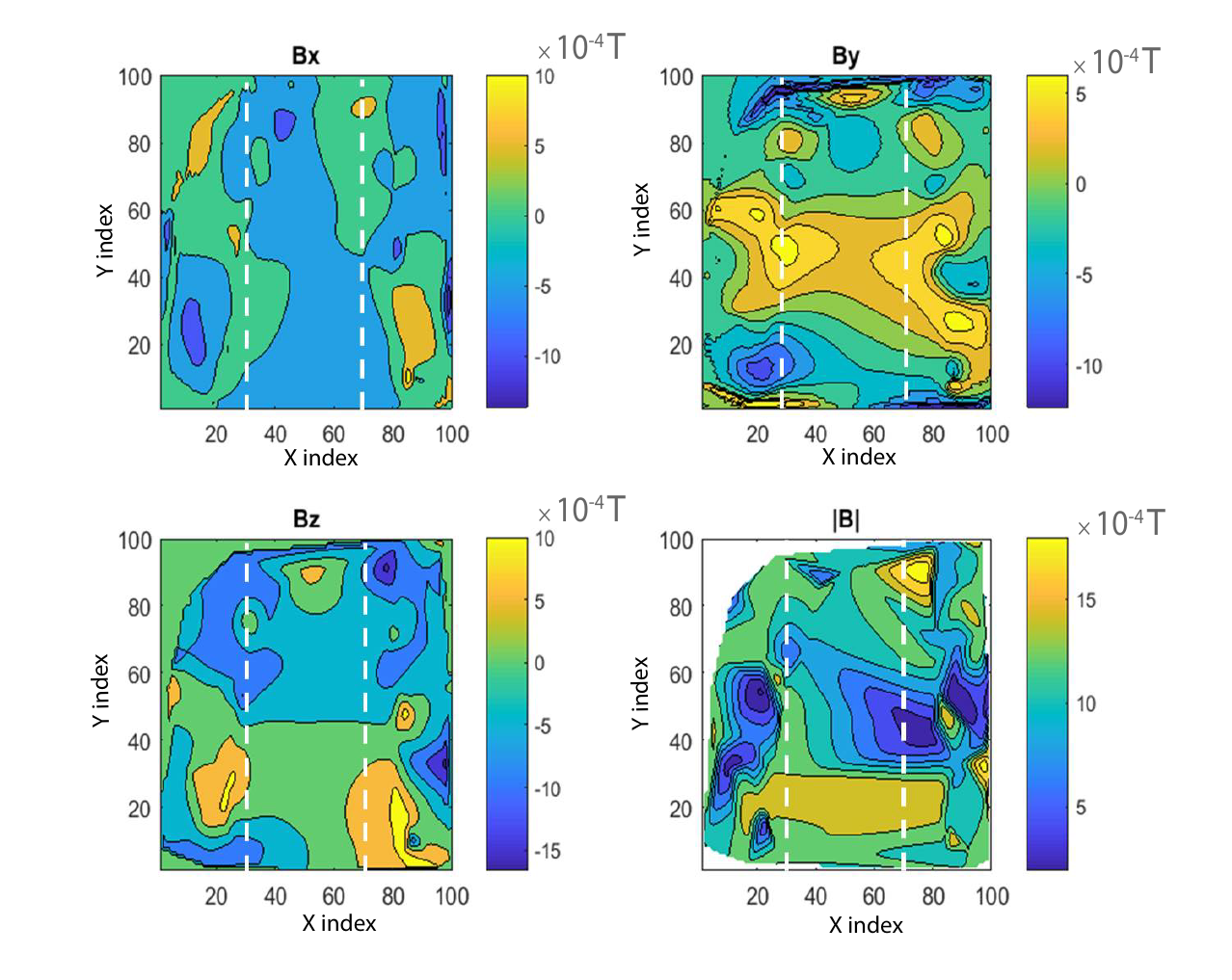}
    \caption{A wide-field magneto-optical image of the magnetic field generated by the GMI microwire using NVs in the 50\,$\mu$m $\times$ 50\,$\mu$m area. $\text{B}_i$ ($i=\text{x,y,z}$) and $|\text{B}|$ are, respectively, the field components along the three orthogonal spatial axes and absolute value of the magnetic field at a given location.  Variation of the field values (as indicated by change in color from yellow to blue) can be attributed to the magnetic field from the different magnetic domains in the GMI wire. An approximate location of the wire is indicated by the white dashed lines.}
    \label{fig_s3}
\end{figure}

The GMI wire had a diameter of $\sim$ 25 $\mu$m, and due to the inherent magnetic nature of the wire its magnetic field extends to NV centers which are about 15-20\,$\mu$m away along the lateral directions (i.e., X and Y-directions the confocal image). Since the wire is magnetic, there exists a spontaneous magnetic field of several Gauss originating from the magnetic domains of the wire at the location of the NV centers. But this field has a spatial profile, in the sense from one point to another along the length of the wire its magnitude and direction change. Figure-\ref{fig_s3} is the wide-field magnetic image of the wire reconstructed using an ensemble of over 70 single NV centers in the given area shown in the confocal image in figure-\ref{fig_s2}. The images were created by recording an ODMR spectra for each of the NV in the confocal image. The ODMR spectra gives the information about the absolute value of the static (bias) field ($\vec{\text{B}}$) experienced by the NV as well as the orientation of the NV axis ($\theta$) with respect to the bias field direction based on the Zeeman relation \(D_{gs}\pm\gamma_{e}.|\text{B}_{||}|\) where $D_{gs}$ is the NV ground state zero field splitting  and $\gamma_{e}$ is the NV gyromagnetic ratio, and $\text{B}_{||}=|\vec{\text{B}}|$cos$\theta$. The components of the magnetic field along the three orthogonal directions as well as the absolute values of the field are then created as shown in the figure-\ref{fig_s3}.\cite{steinert2010high}. The variation in the field values as indicated by both gradual and abrupt changes in the color signifies the presence of different magnetic domains and domain walls in the amorphous GMI wire, their orientation, and their complex interplay. \\
Note that no driving alternating current was applied to the wire when these measurements were done.

\section{\label{sec:calibrate field}Calibrating $B_\text{dc}$ from the coil}

\begin{figure}[h]
    \centering
    \includegraphics[width=\linewidth]{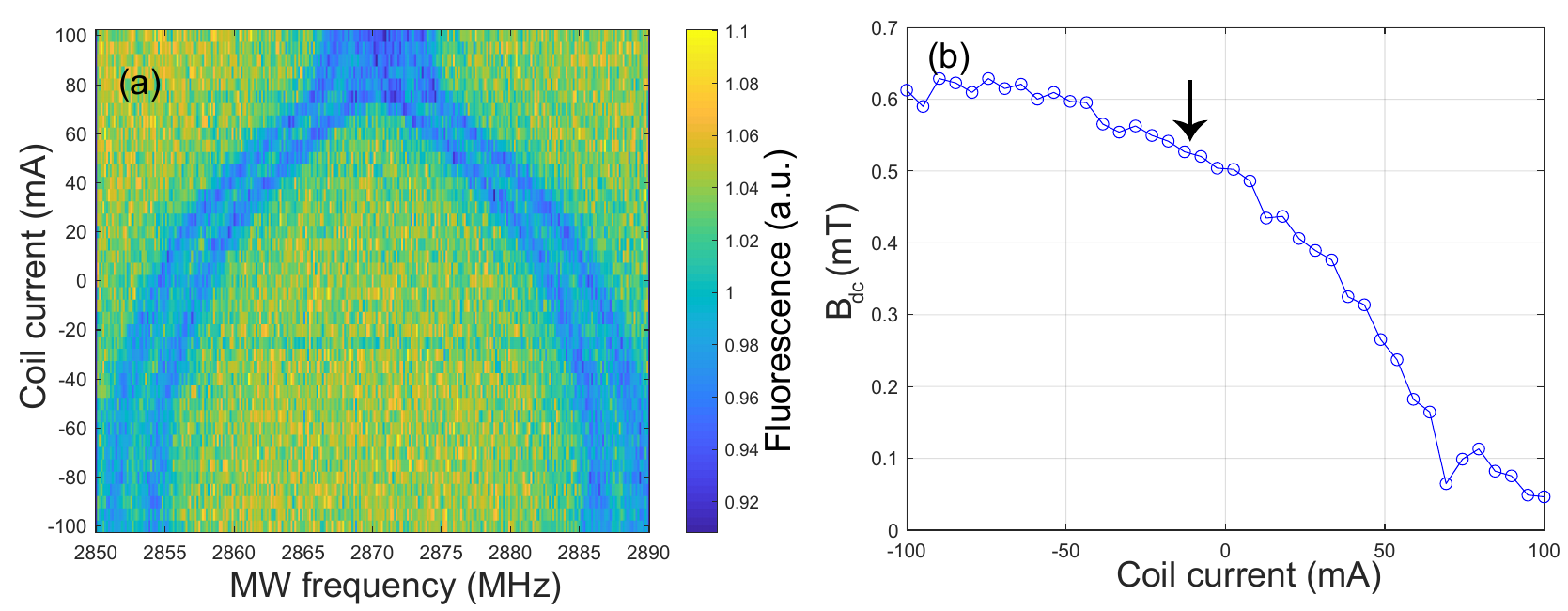}
    \caption{Calibrating the magnetic field generated by current-carrying coil using ODMR. (a) A two-dimensional ODMR spectrum. As mentioned in the figure-2(a) of the main article, the two dips in the ODMR spectra correspond to hyperfine interaction of $^{15}$N nucleus with the NV center. The yellow and blue colors in the color bar indicate the $m_s=\pm1$ and $m_s=0$ sublevels of the NV center, respectively. (b) Extracted values of $B_\text{dc}$ at the location of NV center using the detuned frequency in (a). For demonstrating the magnetometry by sweeping the external field in the main article, most linear part of the plot (shown by black arrow) was chosen.}
    \label{fig_s4}
\end{figure}

As mentioned before, the test magnetic field ($B_\text{dc}$) to be detected was applied through a current-carrying coil in the vicinity (a few cm) of the sample. The amount of magnetic field generated from this coil was estimated by using the standard ODMR experiment. Here, the ODMR spectra was recorded for each current values of the sweepable range of the current source. Figure-\ref{fig_s4}(a) shows such a 2-dimensional scan where the current is swept from -100\,mA to +100\,mA and the corresponding NV fluorescence (indicating the $m_s=0$ population) across the zero field frequency (2870\,MHz) is plotted. The value of $B_\text{dc}$ for the corresponding value of current at the location of NV was then extracted using the Zeeman relation mentioned above (figure-\ref{fig_s4}(b)). For calibrating the $B_\text{dc}$ for magnetometry experiments, we use only a small section in the most linear portion of the data (black arrow in figure-\ref{fig_s4}(b)) and estimate the amount of $B_\text{dc}$ produced over this small current range. We found that the calibration value to be 2.6\,$\mu$T/mA. The calibration constant is only valid for this portion since the slope of the curve changes elsewhere.

We note that the ODMR respose to the current direction (i.e., current sweeping direction from negative to positive or vice versa) was perfectly reversible with a negligible hysteresis.

 The coil was positioned close to the sample and its axis was kept parallel to the GMI wire such that the direction of the DC magnetic field lines emanating from the coil are approximately parallel to the length of the wire. This allows us to take advantage of direction dependent response of the GMI wire to obtain maximum sensitivity for weak external DC fields.

\section{\label{sec:sens analysis}Sensitivity and noise floor analysis}

\begin{figure}[h]

    \centering
    \includegraphics[width=1.0\linewidth]{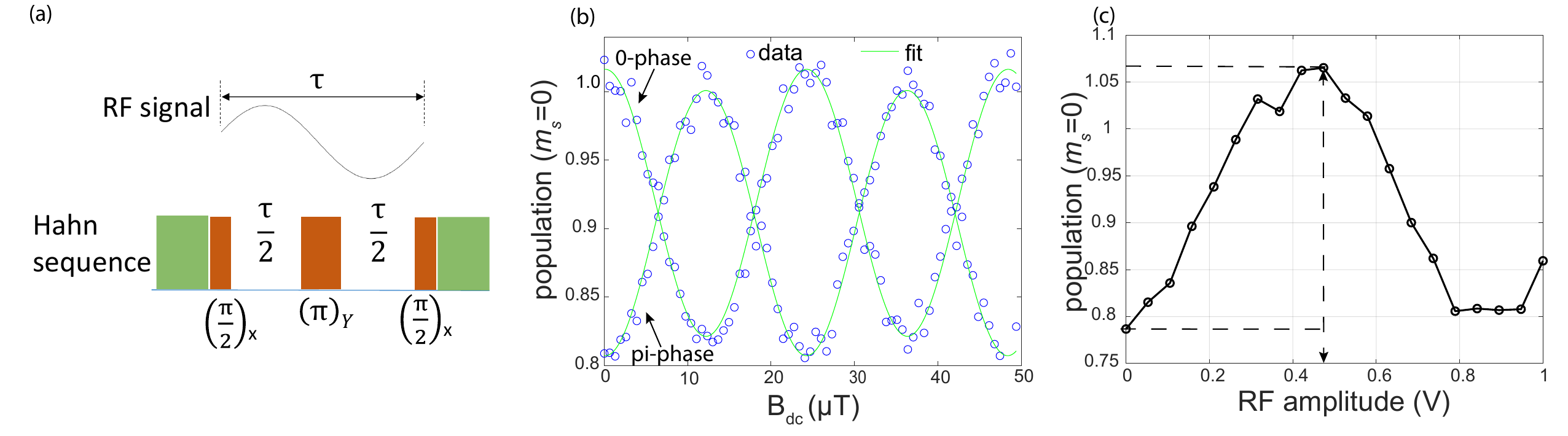}
    \caption{(a) Schematic of the pulse sequence used for sensitivity estimation and noise floor measurement. The sequence is a phase- and frequency-locked Hahn echo protocol referenced to the RF drive signal. (b) A representative magnetometer response showing signal inversion when the phase of the final $\pi/2$ pulse is shifted by $180^\text{o}$. Here $2\tau$ was fixed at 5\,$\mu$s. (c) Measurement of $\Delta B$ generated due to the GMI wire for estimating the magnetometer's noise floor; data were acquired using the protocol in (a), with the RF drive amplitude varied and no signal amplification applied.}
    \label{fig_s6}
\end{figure}

For demonstrating the static magnetic field hybrid sensor based on the GMI wire and a single NV center, we employ the basic scheme presented in the figure-3(a) of the main text (also replicated in the figure-\ref{fig_s6}(a)). Figure-\ref{fig_s6}(b) shows the representative magnetometer response to changes in the external field amplitude. The two datasets represent the same experiment but with the phase of the last (readout) $\pi/2$ pulse differed by 180$^\text{o}$. As explained in the main text, the magnetometer is most sensitive at the highest slope and sensitivity was estimated based on the equation-2 there. The data in figure-\ref{fig_s6}(b) (similar to figure-3(b) of the main text) was fitted with a cosine function, and the error value obtained from the fit is divided by the maximum slope of the signal to obtain the minimum detectable field ($\text{B}_\text{min}$) from which the sensitivity, $\text{B}_\text{min}\sqrt{t}$, was obtained. The measurement time ($t$) used for estimating the sensitivity is given by $t = \sqrt{2 \tau\times n}$, where $2\tau$ is the Hahn-echo sequence time and $n$ is the number of repetitions used for acquiring the data, which is $500,000$ here.

For the noise floor experiment, the Hahn sequence duration was kept constant at $2\tau=10\,\mu$s throughout. To determine the noise floor of our hybrid magnetometer, we use the following procedure. We first estimate the magnetometer contrast and divide it by the magnetic field generated by the wire itself to produce that contrast. This is equivalent to determining the slope in the equation-2 of the main text. For estimating the noise floor of the magnetometer, we acquire the magnetometer signal (i.e., recording the photon count trace) over a period of time 1\,s at a specific sampling frequency (1\,kHz).  We then compute the fast Fourier transform (FFT) of this signal. Dividing the FFT amplitude by the slope mentioned earlier gives the minimum detectable field ($\text{B}_\text{min}$) which is plotted in figure-5 of the main text.

For estimating the contrast, we use the measurement protocol depicted in figure-\ref{fig_s6}(a) for a fixed Hahn duration of $10\,\mu$s and the external test field in the coil being set to 0. We acquire two data sets (not presented here) with the phase of the final $\pi$/2 pulse of the Hahn sequence set 0 and $180^\text{o}$. We then determine the standard deviations, $M_1$ (0-phase) and $M_2$ ($180^\text{o}$-phase), of these two data sets to estimate the magnetometer contrast given by,
\begin{equation}
    \Delta M = \frac{M_1-M_2}{M_1+M_2}
\end{equation}
To estimate the magnetic field (note that this field is different from the intrinsic static field emanating from the magnetic nature of the GMI material) generated by the GMI wire due to driving RF current through it, we again employ the protocol depicted in figure-\ref{fig_s6}(a) with a synchronized RF signal while maintaining zero current in the nearby coil. We slowly ramp the amplitude of the RF voltage in steps and record the fluorescence intensity. The variation of NV fluorescence as a function RF amplitude is shown in figure-\ref{fig_s6}(c). It takes about one 0.5\,V to invert the NV spin from its dark to bright state or about 1\,V to make one full oscillation in the Hahn signal. Since the Hahn sequence's total duration, $2\tau$ was set to 10\,$\mu$s, we can estimate the magnetic field, $\Delta$B, generated over half the oscillation (i.e., in going from dark to bright state) by $\Delta\text{B}=1/(2 \tau \gamma_e)$, where $\gamma_e$ is NV gyromagnetic ratio (28\,GHz/T). So, the ratio $\Delta M/\Delta B$ gives the `slope' of the signal which, along with the FFT amplitude, is then used to estimate the power spectral density showed in figure-\ref{fig5} of the main text.   

\section{\label{sec:Intrinsic GMI noise floor}Intrinsic noise floor of the GMI wire}

The intrinsic magnetic noise spectral density, $\eta_{gmi}$, that defines the theoretically achievable noise floor for the GMI wire sensors is given by\cite{melo2008optimization},

\begin{equation}
    \eta_{gmi} = \frac{2 \alpha k_B T}{\Gamma_e M_s \pi^2 a^2 l},
\end{equation}

where, $\alpha$ is dimensionless damping parameter, $k_B$ is Boltzmann
constant, $T$ is temperature, $\Gamma_e$ is the electron gyromagnetic ratio (28\,GHz/T), $M_s$ is the saturation magnetization value. Using the standard values for $\alpha$ (0.01) and $M_s$ (660\,kA/m) as given in the ref\cite{melo2008optimization} for Co-rich GMI wire (like the one used in our case), we approximate the intrinsic noise floor of our wire to be $\eta_{gmi}\approx 10\, \text{fT}/\sqrt{\text{Hz}}$ at room temperature. Here, we have used nominal values of $a$ and $l$ in our case, 30\,$\mu$m and 10\,mm, respectively. Although the theoretically projected sensitivity is in fT range, however to date, the experimentally reported values for the best noise level are often in the range of $1\,\text{pT}/\sqrt{\text{Hz}}- 100\,\text{pT}/\sqrt{\text{Hz}}$, at least 2-4 orders of magnitude higher than the theoretical value. Many sources contributing to this limitation have been identified \cite{melo2008optimization,ding2009equivalent} with the fundamental limitation being intrinsic magnetic noise associated with the thermal fluctuations of the magnetization of domains in the wire. Other dominant sources include electronic noise in the various circuitry used in the sensor read-out schemes and also from the intrinsic Johnson noise from the GMI wire itself.



%

\end{document}